\documentclass{article}

\usepackage{type1cm}    
\usepackage{makeidx}
\usepackage{graphicx}
\usepackage{hyperref}
\usepackage{multicol}        % used for the two-column index
\usepackage[bottom]{footmisc}% places footnotes at page bottom
\usepackage{amsmath}
\usepackage{newtxtext}       % 
\usepackage[varvw]{newtxmath}       % selects Times Roman as basic font
\usepackage{geometry}
\newcommand\be{\mathbf{e}}

\newcommand\bp{\mathbf{p}}
\newcommand\bq{\mathbf{q}}
\newcommand\bu{\mathbf{u}}

\usepackage{color} 
\newcommand{\delm}[1]{\ifmmode\text{\sout{\ensuremath{#1}}}\else\sout{#1}\fi} % delete math

\begin{document}

% \title*{Modified (hyper-)viscosity for coarse-resolution ocean models}
\title{Modified (hyper-)viscosity for coarse-resolution ocean models}
\author{Louis Thiry, Long Li and Etienne Mémin \\
 INRIA/IRMAR Rennes, France
}
% \institute{Louis Thiry \at INRIA/IRMAR Rennes, France, \email{louis<dot>thiry<at>inria<dot>fr}}

\maketitle

\abstract{We present a simple parameterization for coarse-resolution ocean models.
To replace computationally expensive high-resolution ocean models, we develop a computationally cheap parameterization for coarse-resolution models based solely on the modification of the viscosity term in advection equations.
It is meant to reproduce the mean quantities like pressure, velocity, or vorticity computed from a high-resolution reference solution or using observations.
We test this new parameterization on a double-gyre quasi-geostrophic model in the eddy-permitting regime.
Our results show that the proposed scheme improves significantly the energy statistics and the intrinsic variability on the coarse mesh. 
This method shall serve as a deterministic basis model for coarse-resolution stochastic parameterizations in future works.
}

\section{Introduction}
\label{sec:1}

Ocean general circulation models used at climatic scales are limited for evident computational reasons to too coarse horizontal resolutions to solve correctly ocean mesoscale and sub-mesoscale eddies, even with large computational infrastructure.
The horizontal resolution of the most recent climatic ocean models is of the order of the Rossby radius of deformation.
These models are hence in the so-called eddy-permitting regime and they can solve partially the mesoscale (\textit{i.e.} 10-100 km) eddy field.
These models however suffer from strong limitations.
In particular, they are unable to reproduce accurately large-scale structures such as the eastward turbulent jet in an idealized double-gyre configuration. 

Recent parameterizations have shown significant improvements in coarse-resolution models compared to high-resolution reference solutions \cite{fox2014principles}.
However, it remains an important topic of research, as the actual generation of parametrizations is not completely able to resolve the effects of the unresolved scales on the large-scale flow structures.

A wide range of subgrid parametrizations relies on eddy viscosity such as Laplacian and biharmonic schemes \cite{smagorinsky1963general,leith1996stochastic,griffies2000biharmonic,frisch2008hyperviscosity}.
% Idea of low-pass filter filter which is less dissipative that the laplacian at large-scales.
It has been shown in \cite{jansen2014parameterizing} that including only these (hyper)viscosity in coarse-resolution models often causes too much  dissipation  and results in an  artificial energy sink at large scales. In general, even eddy-permitting models are not energetic enough and as a result, the long-time average of any coarse model's variable of interest departs completely from the long-time average of high-resolution models subsampled at the same scale.
This becomes the main motivation of the present work.
In particular, we would like to answer the following question:
how can we reduce the excessive resolved kinetic energy loss due to the viscosity while simultaneously ensuring numerical stability?

We propose a simple affine parameterization of (hyper)viscosity.
The (bi)laplacian operator $\Delta^p f$ is replaced by $ \Delta^p \left(f - f' \right)$, where $f'$ is a field of same dimension as $f$ that does not depend upon time.
We interpret this method as a mathematical regularization technique to guide the solutions towards prior information.
We frame $f'$ as the solution of an optimal control problem to reproduce statistics computed from a reference solution or observations.
We present a method to solve this optimal control problem.
% It differs from existing \textit{modified hyperviscosity} by \cite{fox2008can} which is a modification of Leith's hyperviscosity \cite{leith1996stochastic} to improve numerical stability.

We test the proposed method with an idealized double-gyre configuration.
For that purpose, we release with this article a fast, concise, and CPU-GPU portable Pytorch implementation of a multi-layer quasi-geostrophic model on a rectangular domain.
We implement and test our optimization procedure within this setting.

This article is organized as follows: we present in section \ref{sec:2} the double gyre quasi-geostrophic model we use and detail its implementation, we present in section \ref{sec:3} our \textit{modified viscosity} parameterization and we show and discuss numerical results in section \ref{sec:4}.

\section{Double gyre quasi-geostrophic model}
\label{sec:2}
\subsection{Governing equations}
We use the same multi-layer quasi-geostrophic model in a non-periodic rectangular domain as in \cite{hogg2005mechanisms}.
Here, we only give a brief review of this system.
The quasi-geostrophic pressure and potential vorticity (PV) are stacked in three isopycnal layers.
We adopt vector forms to denote the layered pressure and potential vorticity (PV):
\[
    \bp =
    \begin{bmatrix}
        p_1\\
        p_2\\
        p_3
    \end{bmatrix}
    , \quad
    \bq =
    \begin{bmatrix}
        q_1\\
        q_2\\
        q_3
    \end{bmatrix} \ .
\]
The forced and damped quasi-geostrophic (QG) equations can be then written as 
\begin{align}
    & \partial_t \bq = \frac{1}{f_0}J (\bq, \bp) + f_0 B \be  + \frac{1}{f_0} \big( a_2 \Delta - a_4 \Delta^2 \big)  (\Delta \bp), \label{adv}\\
    & \big(\Delta - f_0^2 A \big)\bp = f_0 \bq  - f_0 \beta (y - y_0) , \label{elliptic}
\end{align}
where $\Delta = \partial_{xx}^2 + \partial_{yy}^2$ denotes the horizontal Laplacian, $\Delta^2$ the bi-laplacian operator, $J (a, b) = \partial_x a \partial_y b - \partial_x b \partial_y a$ stands for the Jacobi operator,
$f_0 + \beta (y - y_0)$ is the Coriolis parameter under beta-plane approximation with the meridional axis center $y_0$, $a_2$ and $a_4$ are the Laplacian and biharmonic viscosity coefficients.
Parameters of the configuration are listed in the Tables \ref{tab:params-pub} and \ref{tab:params-pri} in Appendix.
Besides, the second term on the right-hand side of equation (\eqref{adv}) represents the external forcing applied on different layers. In this work, we only consider an idealized case in which the ocean basin is driven by a stationary and symmetric wind stress $\vec{\tau} = (\tau^x, \tau^y)$ on the surface and by a linear Ekman stress at the bottom. In that case, the forcing term can be specified by
\[
    B =
    \begin{bmatrix}
    \frac{1}{H_1} & \frac{-1}{H_1} & 0 & 0 \\
    0 & \frac{1}{H_2}  &  \frac{-1}{H_2} & 0 \\
    0 & 0 &\frac{1}{H_3} & \frac{-1}{H_3}\\
    \end{bmatrix},\ \quad 
    \be =
    \begin{bmatrix}
    \partial_x \tau^y - \partial_y \tau^x\\
    0\\
    0\\
    \frac{\delta_{\rm ek}}{2 |f_0|} \Delta p_3
    \end{bmatrix},\ \quad
    \vec{\tau} = \tau_0
    \begin{bmatrix}
     -\cos(2 \pi y / L_y)\\
     0
    \end{bmatrix},
\]
where $\tau_0$ is the magnitude of surface wind, $H_k$ is the background thickness of layer $k$, and $\delta_{\rm ek}$ is the bottom Ekman layer thickness. The vertical stratification level of such a model is described by the term $- f_0^2 A \bp$ in Equation \eqref{elliptic} with %the matrix $A$ defined by
\[ A = 
    \begin{bmatrix}
    \frac{1}{H_1 g_{1.5}'} & \frac{-1}{H_1 g_{1.5}'} & 0\\
    \frac{-1}{H_2 g_{1.5}'} & \frac{1}{H_2}\left( \frac{1}{g_{1.5}'} + \frac{1}{g_{2.5}'}\right)  &  \frac{-1}{H_2 g_{2.5}'}\\
    0 &\frac{-1}{H_3 g_{2.5}'} & \frac{1}{H_3 g_{2.5}'}
    \end{bmatrix}
    ,
\]
where $g_{k+0.5}$ is the reduced gravity defined across the interface between layers $k$ and $k+1$.
A multi-layered generalization of this model can be found in \cite{hogg2003quasi}.
Note also that such a multi-layered model can be considered as a vertical discretized approximation of the continuously stratified QG system \cite{Vallis2017atmos} with $\partial_z (f_0 \partial_z \bp / \vec{N}^2) \approx -f_0 A \bp$ approximated by finite differences, and in which $\vec{N}$ denotes the buoyancy (or Brunt-Vaisala) frequency.

\subsection{Pytorch implementation}

To facilitate numerical developments and benefit from built-in automatic differentiation, we develop a Pytorch \cite{paszke2019pytorch} implementation of the above-described multilayer QG model \footnote{Available at https://github.com/louity/qgm\_pytorch}.
For this purpose, we follow rigorously the strategy of \cite{hogg2014formulation}:
\begin{enumerate}
    \item we use a regular numerical grid with finite differences
    \item We solve the PV advection equation \eqref{adv} on the whole domain except the boundaries.
    We use a standard 5-point finite difference scheme for the (bi-)laplacian and the energy-enstrophy conservative Arakawa-Lamb scheme for the Jacobian \cite{arakawa1981potential}.
    \item We apply a vertical change of coordinate to equation \eqref{elliptic} which  becomes a set of three inhomogeneous Helmholtz equations.
    We solve these equations with the spectral Discrete Sine Transform (DST) method, and we add corresponding homogeneous Helmholtz equation solutions to ensure mass conservation.
    \item We update the boundary values of the potential vorticity $\bq$ using equation \eqref{elliptic}.
\end{enumerate}
Detailed equations and numerical routine design choices can be found in \cite{hogg2014formulation}.
We use a Heun-Runge-Kutta 2 time-stepping instead of the Leap-Frog time scheme used by \cite{hogg2014formulation}.

For sake of numerical efficiency, we follow the recommendation of \cite{roullet2021fast}: we compile computationally demanding routines and simplify finite difference calculations by reducing as much as possible the number of multiplications.
We end up with a very concise code (less than 300 lines) that only depends upon Numpy and Pytorch libraries.
% Ensembles of $N$ initial conditions can be easily integrated in time in parallel with this code.
% Speed performance for the advection equation \ref{adv} are close to performances of \cite{roullet2021fast}, \textit{i.e.} close to maximal arithmetic intensity on both Intel CPU and Nvidia GPU.
This implementation will be open-sourced at the time of the publication.

\subsection{Eddy-resolving and eddy-permitting regimes}

We consider two spatial settings for our simulations :
\begin{enumerate}
    \item The eddy-resolving regime, our high-resolution reference with a 5 km resolution.
    \item The eddy-permitting regime, our low-resolution setting with a 40 km resolution.
\end{enumerate}
Parameters for these two different regimes are written in table \ref{tab:params-pri} in Appendix.

\begin{figure}
    \centering
    \includegraphics[width=0.99\linewidth]{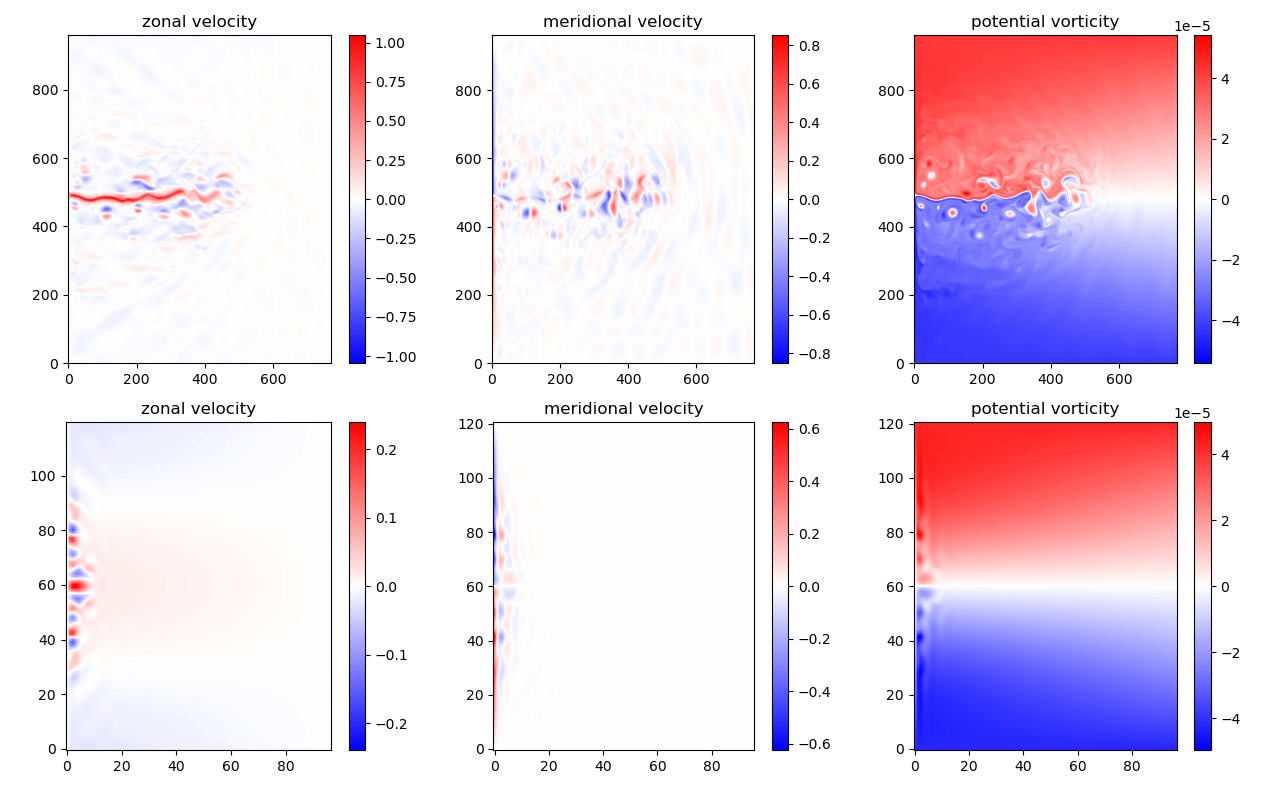}
    \caption{(Top) high-resolution and (bottom) low-resolution top-layer snapshots after 400 years of integration starting from zero velocity. Velocities are in m s$^-1$ and PV in s$^{-1}$.}
    \label{fig:snapshots}
\end{figure}

\cite{shevchenko2015multi} studied the resulting flows' differences between these two regimes.
The high-resolution eddy-resolving model shows a well-pronounced eastward jet fuelled by mesoscale eddies circulating while the low-resolution eddy-permitting model does not induce a proper eastward jet as shown on Fig. \ref{fig:snapshots}.
Temporal statistics significantly differ between  high- and low-resolution simulations.

\section{Proposed modified viscosity}
\label{sec:3}
\subsection{Motivation}

\begin{figure}
    \centering
    \includegraphics[width=0.99\linewidth]{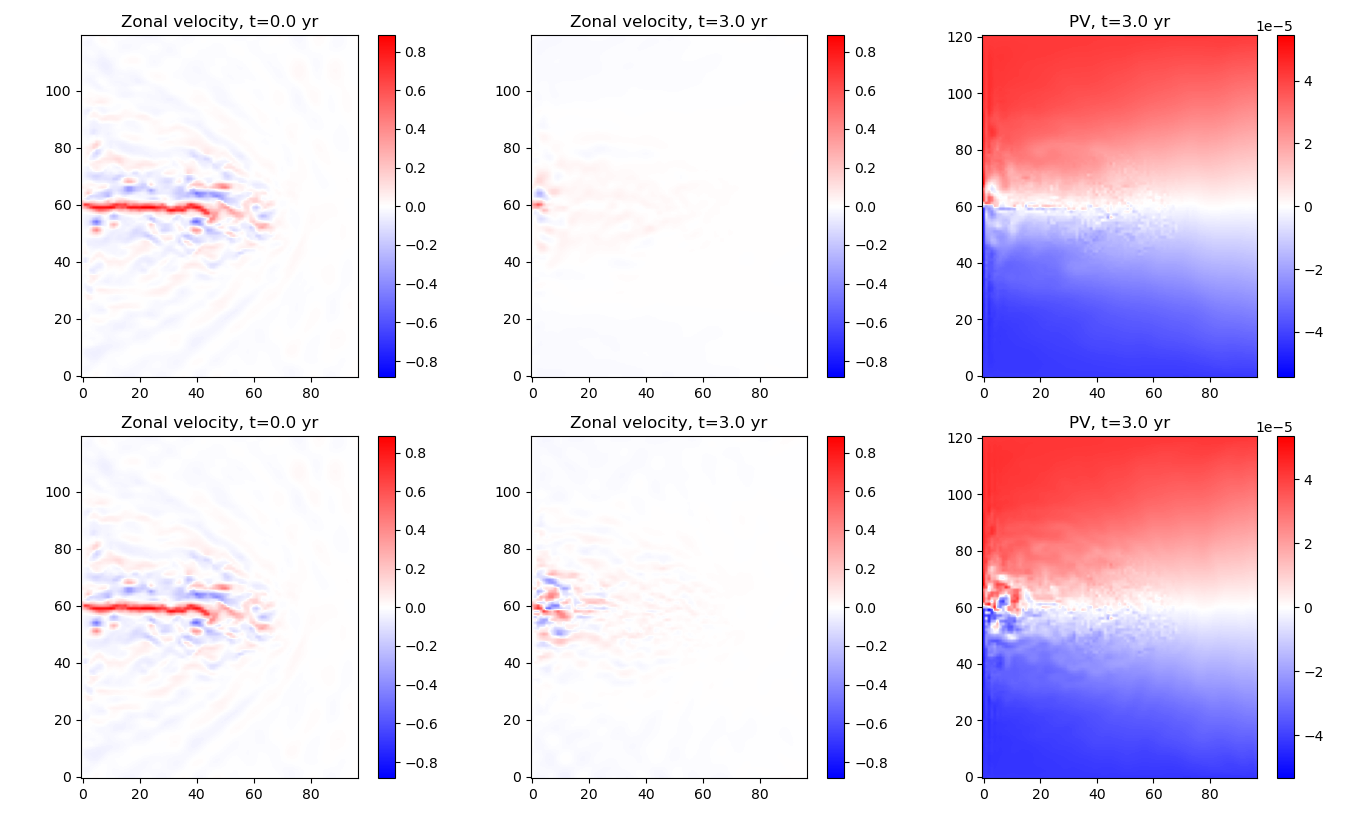}
    \caption{(left) Initial condition: high-resolution snapshot on the low-resolution grid.(center and right) Zonal velocity and potential vorticity (PV) snapshots after 3 years of integration at low-resolution with equations (\ref{adv},\ref{elliptic}) with (top) standard hyper-viscosity and (bottom) 10 times smaller hyper-viscosity.
    We can see aliasing effects on potential vorticity snapshots integrated with low hyper-viscosity.
    }
    \label{fig:2}
\end{figure}

In both resolutions, we use biharmonic viscosity as in \cite{smagorinsky1963general,leith1996stochastic,griffies2000biharmonic,frisch2008hyperviscosity} essentially because it is less dissipating at large scales than a Laplacian.
Compared to the usual Laplacian viscosity, it preserves large-scale structures.
However, hyperviscosity remains much too dissipative in  the “eddy-permitting” regime \cite{jansen2014parameterizing}.
This too strong dissipation \textit{kills} the eastward jet that is present in the high-resolution and that we expect to see in such a double-gyre quasi-geostrophic model.
Fig. \ref{fig:2} shows a sequence of snapshots of the low-resolution models where we input a downsampled snapshot of the high-resolution (see Appendix for details on downsampling).
After as few as three years, the eastward jet has almost disappeared, showing that the model is too dissipating.
Lowering the hyper-viscosity coefficient by a factor of 10 does not solve this problem, and creates spurious gradients in the potential vorticity as shown in Fig. \ref{fig:2}.
These numerical artifacts are due to a bad representation of the direct enstrophy cascade,  causing a piling up of the small-scale vorticity gradients at the cut-off frequency together with aliasing effects.

\subsection{Modified viscosity}

Here we propose a simple affine modification parameterization of hyperviscosity.
We add a bias to the term $\Delta \bp $ in eq. \eqref{adv}, which becomes $ \Delta \left(\bp - \bp' \right) $ where $\bp' $ is a  dimensional field that does not depend upon time.
The PV advection equation with hyperviscosity becomes
\begin{align}
    & \partial_t \bq = \frac{1}{f_0}J (\bq, \bp) + f_0 B \be  + \frac{1}{f_0} \big( a_2 \Delta - a_4 \Delta^2 \big)  \left(\Delta (\bp - \bp')\right).\label{adv_m}
\end{align}
The elliptic equation \eqref{elliptic} remains unchanged.

The goal of this additional term is to reproduce a relevant time-average pressure field relying on observations or high-resolution solutions.
For example the high-resolution average $\overline{\bp_{\rm HR}}$ can be downsampled to the targeted coarse grid  resolution in $\overline{\bp_{\rm HR}}\downarrow$, and we want the average of the modified low-resolution $\overline{\bp_{\rm LR}}$ model to be as close as possible to the high-resolution reference $\overline{\bp_{\rm HR}}\downarrow$.

We face here an optimal control problem, as the low-resolution average is a function of the control parameter $\bp'$.
We state it with the following least-square formulation
\begin{align}
    \bp'_{\rm opt} &= \underset{\bp'}{\text{argmin }} \mathcal{F}\left( \bp'\right) \label{opt_ctrl}\\
    \mathcal{F}\left( \bp'\right)&=  \big\| \ \overline{\bp_{\rm LR}}\left( \bp'\right) - \overline{\bp_{\rm HR}}\downarrow \big\|^2 
\end{align}

This optimization problem is \textit{a priori} non-convex and we shall not expect to find a global optimum.
In the following, we propose a numerical procedure to find a heuristic $\hat{\bp'}$ of the optimal solution $\bp'_{\rm opt}$.

Computationally, the implementation of this modified hyperviscosity is simple and computationally cheap.
We precompute $\Delta \bp'$ and subtract it from $\Delta \bp$ at each time-integration step.
It increases the integration time of the advection equation \eqref{adv} by less than  1\% on CPUs and GPUs.

\subsection{Modified viscosity regularization}

The continuously stratified  QG equations can be rewritten in a variational formulation \cite{holm1998hamilton} with a Hamiltonian $\mathcal{J}$ defined as 
\[
\mathcal{J} (\bp) = \frac{1}{2} \int_\Omega \frac{1}{f_0} |\nabla \bp|^2 + \frac{f_0}{N^2} (\partial_z \bp)^2 .
\]
Our model is a discretized version of the continuous stratification.
Since we add an external wind forcing term and we use an energy conservative Arakawa advection scheme, we need to add some viscosity or hyperviscosity to dissipate energy.
In a variational formulation, these (hyper-)viscous terms become
the following penalization
\[ \frac{1}{2}\int_{\Omega} a_2 |\Delta\bp|^2 + a_4 \left| \nabla\left( \Delta \bp \right) \right|^2,
\]
added to the Hamiltonian $\mathcal{J} (\bp)$ to produce a smooth solution.
The Gradient norm penalization of Laplacian $\bp$ guides the minimization toward solutions of smooth Laplacian. 
Hyperviscosity corresponds to the Laplacian norm penalization and enforces a solution of minimum Laplacian norm. The parameters $a_2$ and $a_4$ quantify the strength of these regularization constraints.

Here, we simply propose to replace it with the following penalization 
\[\frac{1}{2}\int_{\Omega} a_2 |\Delta (\bp - \bp')|^2 + a_4 \left| \nabla \left(\Delta (\bp - \bp')\right)\right|^2 \ .
\]
We now penalize $(\bp - \bp')$ instead of $\bp$, meaning that we guide the solution to a possibly non-smooth reference $\bp'$ that will produce the correct large scale behavior.

\subsection{Iterative procedure}

Here we present a method to find a solution to the optimization problem \eqref{opt_ctrl}.
A natural guess for $\bp'_{\rm opt}$ is $\overline{\bp_{\rm HR}}\downarrow$.
We solve the equations and compute the average pressure $\overline{\bp_{\rm LR}}$.
Results are shown in Fig. \ref{fig:4}. It is a good first-guess, but the difference $\overline{\bp_{\rm HR}}\downarrow - \overline{\bp_{\rm LR}} $ is still large.

We propose the following iterative procedure to find a better guess for $\bp'_{\rm opt}$. 
In the following we assume that we are in low resolution, \textit{i.e.} $\bp = \bp_{\rm LR}$ and $\overline{\bp} = \overline{\bp_{\rm LR}}$ unless explicitly written.
\begin{itemize}
    \item We set  $\bp'_0$ and we compute the average pressure $\overline{\bp}_0$ solving standard equations (\ref{adv},\ref{elliptic}) without modified viscosity.
    \item Choose $k \in \left] 0, 1 \right]$.
    \item Start with $\bp'_1 = \overline{\bp_{\rm HR}}\downarrow$.
    \item Evolve the ensemble for $n$ years and compute the corresponding average pressure $\overline{\bp}_1$ with ensemble average.
    % \item While $\| \overline{\bp}_{n} - \overline{\bp}_{n-1}\|^2 >   \sigma^2 \left( \|\overline{\bp}_{n}\|\right) / N$:
    \item For $n = 1 \dots$:
    \begin{itemize}
        \item Set $\bp'_{n+1} = \bp'_{n} + k \left( \overline{\bp_{\rm HR}}\downarrow  - \overline{\bp}_n\right)$.
        \item  Evolve the ensemble for $n$ years and compute new average pressure $\overline{\bp}_{n+1}$.
    \end{itemize}
    \item return $\bp'_{n}$ and $\overline{\bp}_{n}$
\end{itemize}

% In the convergence criterion
% \[ \| \overline{\bp}_{n} - \overline{\bp}_{n-1}\|^2 >   \sigma^2 \left( \|\overline{\bp}_{n}\|\right) / N \]
% the term $\sigma^2 \left( \|\overline{\bp}_{n}\|\right) / N $ corresponds to the variance of the statistical fluctuations of the norm of the average pressure $\|\overline{\bp}_{n}\|$.
% If the difference between two consecutive averages is smaller than this quantity, it can be due only to statistical fluctuations.
% Hence we estimate that the procedure has converged.
There is no theoretical guarantee that this procedure converges, but we observe in the next section that it converges with the double-gyre QG model that we use.

% \subsection{Auto-differentiation gradient descent}

% To optimize $\bp'$, we can leverage the Pytorch automatic differentiation (autodiff) to solve the optimization problem \ref{opt_ctrl}.
% Pytorch's autodiff is based on the backpropagation algorithm which requires storing all intermediate steps in memory.
% With our memory constraints, we can not integrate with autodiff for 10 years, minimum time required to have converged statistics \cite{ryzhov2020data}.
% Hence we propose the following procedure:
% \begin{itemize}
%     \item Start from $\bp'_0 = \bp'_n$ obtained with the iterative procedure mentioned below.
%     \item While $\| \overline{\bp}_{n} - \overline{\bp}_{n-1}\|^2 >   \sigma^2 \left( \|\overline{\bp}_{n}\|\right) / N$:
%     \begin{itemize}
%         \item Turn-off automatic differentiation and spin up the model for 10 to 100 years to have converged statisics.
%         \item Turn-on automatic differentiation and run the model as long as possible given the limited memory constraints.
%         \item Compute the average pressure $\overline{\bp}_{n}(\bp'_n)$.
%         \item Perform a gradient descent step:
%         \[ \bp'_{n+1} = \bp'_{n} - \delta \nabla_{\bp'_{n}}\mathcal{F}(\bp'_n) \]
%     \end{itemize}
%     \item return $\bp'_{n}$ and $\overline{\bp}_{n}$
% \end{itemize}

\section{Results and discussion}
\label{sec:4}

\subsection{Statistics}

We use ensemble averages to compute the statistics.
To create ensembles of size $N$, we start from a zero solution and spin up the models for 100 years with a timestep of 1200 seconds to reach statistically steady states as in \cite{ryzhov2020data}.
Then we run the models for 500 years and save 10 snapshots a year to get 5000 snapshots, and we randomly select $N$ snapshots out of these 5000 snapshots.
The ensemble averages are simply average over these $N$ ensemble members that we evolve in parallel.
Such ensemble averages are denoted with $\overline{\bullet}$ in the following, \textit{i.e.} the average pressure is denoted by $\overline{\bp}$, average velocity by $\overline{u}$, etc.

\subsection{Iterative procedure}
\label{ssec:it_proc}

\begin{figure}
    \centering
    \includegraphics[width=0.95\linewidth]{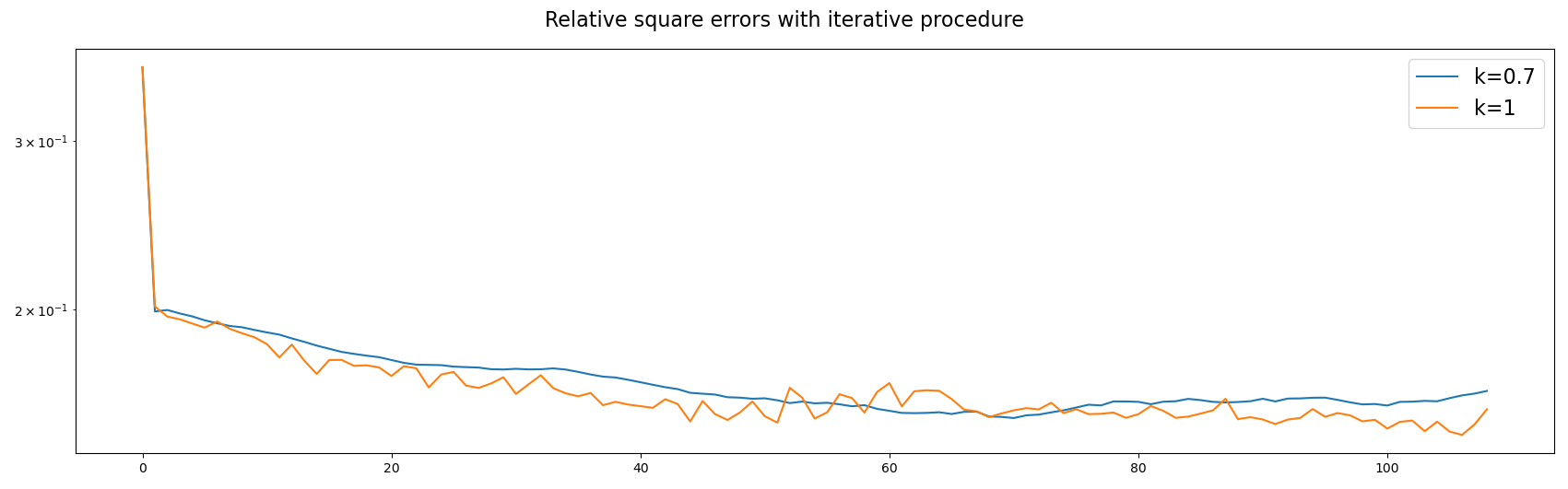}
    \caption{Evolution of the relative square error $ \frac{\| \overline{\bp}_{n} - \overline{\bp_{HR}}\downarrow\|^2}{ \|\overline{\bp_{HR}}\downarrow\|^2}$ w.r.t iterations of the procedure.}
    \label{fig:3}
\end{figure}

We test the iterative procedure described in section \ref{ssec:it_proc} with the double-gyre model presented in section \ref{sec:2} in the eddy-permitting regime.
We use $n=10$ years to evolve the ensemble after each iterate.
We compute the reference pressure average $\overline{\bp}_{\rm HR}$ with the same model in the eddy-resolving regime.

\begin{figure}
    \centering
    \includegraphics[width=0.99\linewidth]{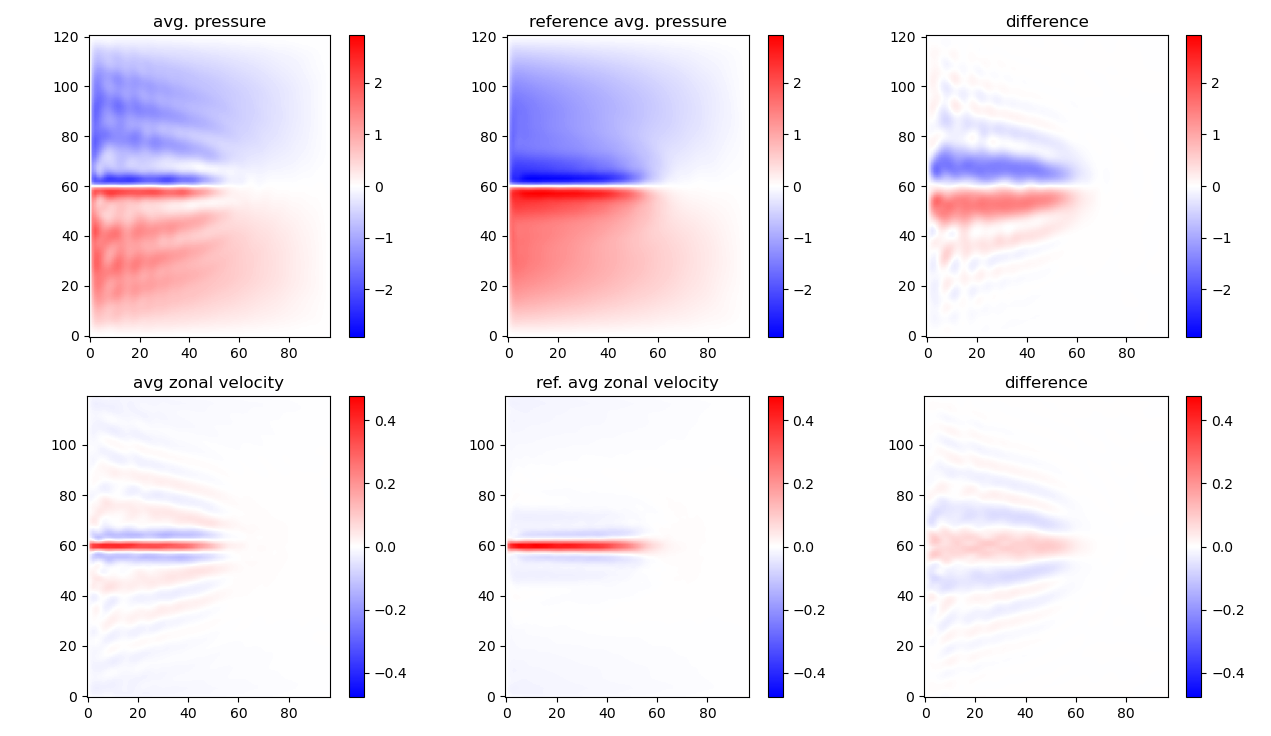}
    \caption{Top-layer average pressure (top) and velocity (bottom) of (left-to-right) proposed model at low-resolution, reference, and the difference between the two.
    }
    \label{fig:4}
\end{figure}

Fig. \ref{fig:3} shows the relative square error $ \| \overline{\bp}_{n} - \overline{\bp_{HR}}\downarrow\|^2 / \|\overline{\bp_{HR}}\downarrow\|^2$ at iterations of the procedure with $k=1$ and $k=0.7$.
The procedure converges with $k=0.7$ and oscillates with $k=1$.

\begin{figure}
    \centering
    \includegraphics[width=0.95\linewidth]{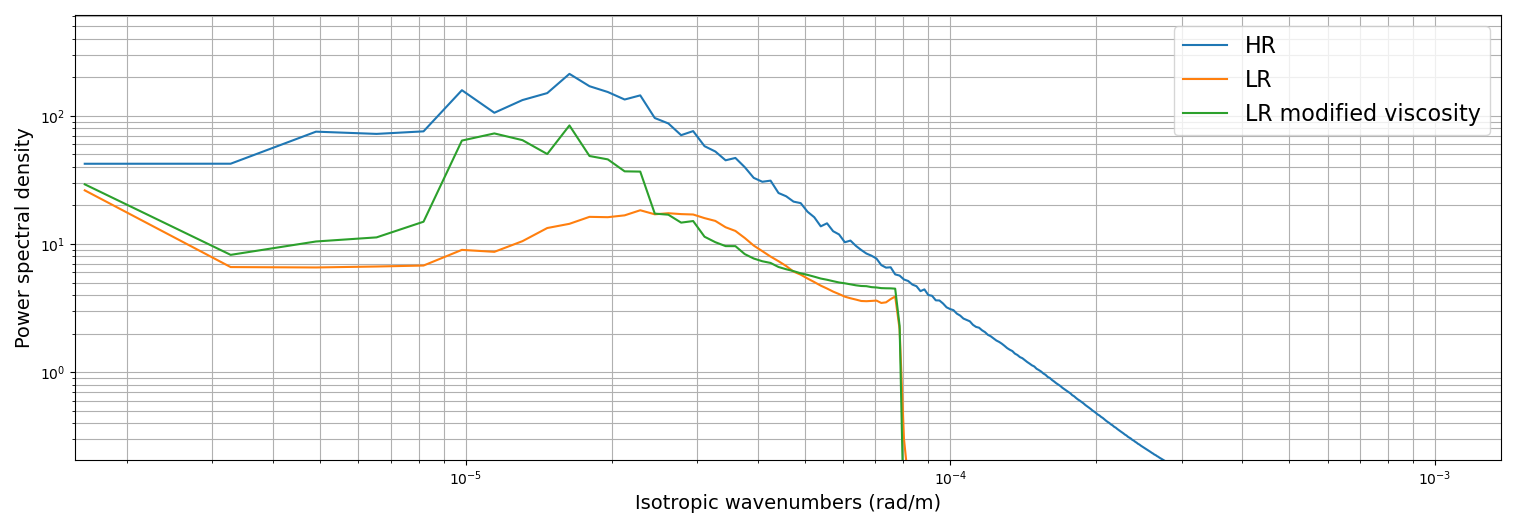}
    \caption{Top-layer kinetic-energy spectra average with models at high-resolution (HR), at low-resolution (LR) and at low-resolution with proposed modified viscosity.
    The decreasing slope of the spectrum of the proposed model is much closer to the high-resolution reference.}
    \label{fig:5}
\end{figure}

\begin{figure}
    \centering
    \includegraphics[width=0.99\linewidth]{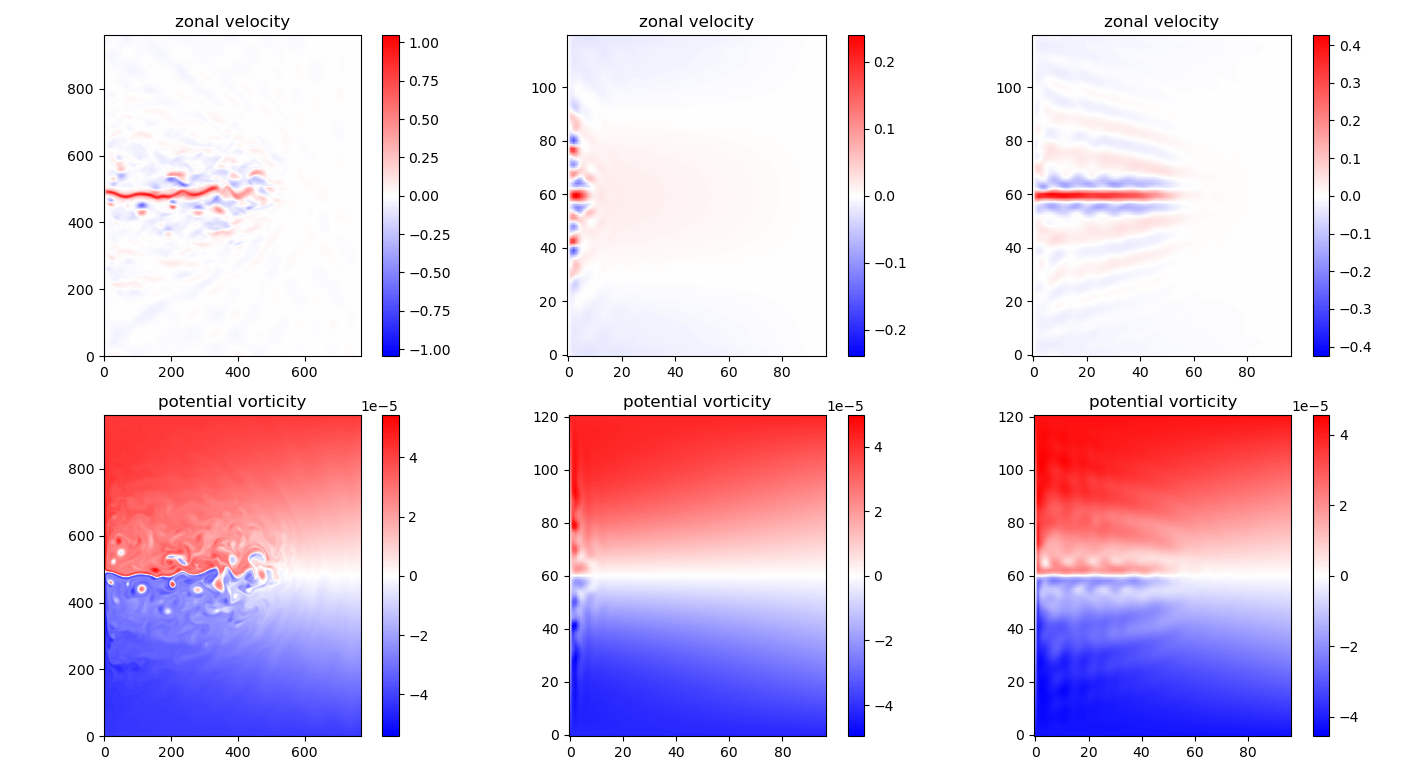}
    \caption{PV and zonal velocity snapshots form (left-to-right)  high-resolution, low-resolution and proposed model at low-resolution.
    }
    \label{fig:6}
\end{figure}

Fig. \ref{fig:4} shows the output average pressure $\overline{\bp}_{n}$ of the iterative procedure, the reference $\overline{\bp}_{\rm HR}$ and the difference between the two, as well as for zonal velocity $\bu$ .
Our model can reproduce the eastward jet produced by the high-resolution reference model.
Kinetic energy spectra shown on Fig. \ref{fig:5} shows also the improvement of our model compared to low-resolution.
Finally, Fig. \ref{fig:6} shows high-resolution and low-resolution snapshots as well as a snapshot of the proposed model at low-resolution.
Our model effectively produces the eastward jet and a re-circulation zone around it where eddies are created.
Artifacts can be also observed on the zonal velocity and potential vorticity on the right of Fig. \ref{fig:6}.
They can likely be Rossby waves created by the harmonic regularization terms, which remain an artificial constraint, but this needs to be studied further.  

\section{Conclusion}
\label{sec:5}
We presented a simple modified-viscosity scheme for coarse resolution ocean modeling that we derived and tested on a double-gyre multi-layer quasi-geostrophic model.
We interpret it as a modified regularization technique that will guide the solution to a reference rather than producing a too smooth solution in the eddy-permitting regime.
The technique requires solving an optimization problem, and we presented a procedure to find a good guess for the solutions.
We showed that it converges to a reasonable solution that fairly reproduces the input reference.

If this method mimics the average of the high-resolution, it only reproduces partially the variability and higher-order statistics of the high-resolution.
We see in Fig. \ref{fig:5} our model's snapshots resemble the averages.
In future works, we consider using this method as a deterministic basis for stochastic parameterizations such as Location-Uncertainty \cite{memin2014fluid}.

\clearpage
\section*{Acknowledgments}
The authors acknowledge the support of the ERC EU project 856408-STUOD. 
\section*{Appendix}
\addcontentsline{toc}{section}{Appendix}

\subsection*{Downsampling procedure}

Downsampling the high-resolution solution on a low-resolution grid consists of interpolating the high-resolution (769$\times$ 961) streamfunction on the low-resolution (97$\times$121) grid.
Then we can compute the potential vorticity using equation (\ref{elliptic}).
Because of the no-flow constraint, the downsampled  streamfunction should be constant on the boundaries and should satisfy a mass conservation constraint \cite{hogg2014formulation}.
We also want to preserve the frequency information and prevent aliasing.

We use the following procedure:
\begin{enumerate}
    \item we apply a Gaussian filter and downsample the streamfunction on the domain except on the boundaries.
    \item we adding homogeneous solutions of the Elliptic equation (\ref{elliptic}) to the streamfunction in order to satisfy the mass conservation as in \cite{hogg2014formulation}.
\end{enumerate}

\subsection*{Parameter tables}
\begin{table}[htbp]
\begin{center}
%\resizebox{0.8\textwidth}{!}{%
\begin{tabular}{ccc}
\hline
Parameters                     & Value                                                  & Description                      \\ \hline
$L_x \times L_y$                   & $(3840\times 4800)$ km                    & Domain size               \\
$H_k$                             & $(350, 750, 2900)$ m                       & Mean layer thickness   \\
$g_k'$                             & $(0.025, 0.0125)$ m s$^{-2}$              & Reduced grativity       \\
$\delta_{\text{ek}}$         & $2$ m                                                & Bottom Ekman layer thickness   \\
$\tau_0$                         & $2\times 10^{-5}$ m$^2$ s$^{-2}$           & Wind stress magantitude \\ 
$a_2$                             & $0$ m$^2$ s$^{-1}$                                 & Laplacian viscosity coefficient               \\   
$f_0$                              & $9.375\times 10^{-5}$ s$^{-1}$            & Mean Coriolis parameter   \\
$\beta$                           & $1.754\times 10^{-11}$ (m s)$^{-1}$    & Coriolis parameter gradient   \\
% $L_d$                             & $(39, 22)$ km                                   & Baroclinic Rossby radii   \\
\hline
\end{tabular}
%}
\caption{Common parameters for all the models.}
\label{tab:params-pub}
\end{center}
\end{table}

\begin{table}[htbp]
\begin{center}
%\resizebox{0.7\textwidth}{!}{%
\begin{tabular}{cccc}
\hline
Grid dimensions & Resolution       & Timestep      & Hyperviscosity ($a_4$) \\ \hline
769 $\times$ 961 & $5$ km                                           & $600$ s                                 & $2.0\times 10^{9}$ m$^4$ s$^{-1}$              \\
97 $\times$ 121 & $40$ km                                         & $1200$ s                                 & $5.0\times 10^{11}$ m$^4$ s$^{-1}$               \\
%$80$ km                                         & $1440$ s                                & $5.0\times 10^{12}$ m$^4$ s$^{-1}$               \\
%$120$ km                                       & $1800$ s                                 & $1.0\times 10^{13}$ m$^4$ s$^{-1}$               \\
\hline
\end{tabular}
%}
\caption{Grid-dependent parameters.}
\label{tab:params-pri}
\end{center}
\end{table}

\clearpage

\end{document}